# Atomic-scale imaging of emergent order at a magnetic field–induced Lifshitz transition

Carolina A. Marques[1]†, Luke C. Rhodes[1]†, Izidor Benedičič[1], Masahiro Naritsuka[1], Aaron B. Naden[2], Zhiwei Li[3]‡, Alexander C. Komarek[3], Andrew P. Mackenzie[1,3], Peter Wahl[1]*

The phenomenology and radical changes seen in material properties traversing a quantum phase transition have captivated condensed matter research over the past decades. Strong electronic correlations lead to exotic electronic ground states, including magnetic order, nematicity, and unconventional superconductivity. Providing a microscopic model for these requires detailed knowledge of the electronic structure in the vicinity of the Fermi energy, promising a complete understanding of the physics of the quantum critical point. Here, we demonstrate such a measurement at the surface of $Sr_3Ru_2O_7$. Our results show that, even in zero field, the electronic structure is strongly $C_2$ symmetric and that a magnetic field drives a Lifshitz transition and induces a charge-stripe order. We track the changes of the electronic structure as a function of field via quasiparticle interference imaging at ultralow temperatures. Our results provide a complete microscopic picture of the field-induced changes of the electronic structure across the Lifshitz transition.

## INTRODUCTION

In strongly correlated electron materials, electron interactions lead to a plethora of ground states in close vicinity to each other, making the materials highly sensitive to external stimuli and thus attractive for technological applications. However, in most cases, a microscopic understanding of these phases is missing because of a lack of knowledge about the electronic structure in the vicinity of the Fermi energy. In many cases, the most interesting region of the phase diagram is in the vicinity of a quantum critical point, giving rise to a phase transition which is driven by quantum fluctuations rather than thermal fluctuations. Among the exotic phases observed near quantum critical points are nematic electronic states (*1*) and unconventional superconducting phases (*2*) including high-temperature superconductivity (*3*). To provide a full understanding of the physics of a quantum phase transition, it is highly desirable to be able to track the electronic structure spectroscopically as a material traverses such a transition. One of the cleanest systems in which a field-tuned metamagnetic transition has been interpreted in terms of quantum criticality is $Sr_3Ru_2O_7$ (*1*, *4*), although it remains controversial whether the physics of the material is driven by quantum fluctuations or can largely be accounted for by van Hove singularities (vHss) in the single-particle density of states in the vicinity of the Fermi energy (*5*–*9*). The bulk properties of the bilayer member of the Ruddelsden-Popper series of the strontium ruthenates (Fig. 1A) feature a series of metamagnetic transitions around a magnetic field of 8 T, accompanied by the formation of magnetically ordered phases (*10*). Traces of the putative quantum critical behavior are observed in a wide temperature range in the temperature dependence of various physical properties including resistivity (Fig. 1B) (*4*), specific heat (*5*, *7*), magnetostriction (*11*), and magnetization (*12*). It is generally accepted that van Hove singularities in the vicinity of the Fermi energy can play an important role for the physical observables close to a quantum critical point (*9*, *13*, *14*), yet detailed knowledge of the electronic structure of $Sr_3Ru_2O_7$ on the relevant energy scales is missing, which would place constraints on models of its quantum criticality, metamagnetism, and the nematic state (*9*, *15*–*17*). Establishing a microscopic model requires detailed knowledge of the electronic structure on the relevant energy scales of a few hundred micro–electron volts around the Fermi energy, as can be obtained in principle by quasiparticle interference imaging (QPI) (*18*). Here, we use quasiparticle interference imaging to unravel the surface electronic structure of $Sr_3Ru_2O_7$ with sub–milli–electron volt resolution. Ultralow-temperature scanning tunneling microscopy (STM) at temperatures below 100 mK allows us to determine the field-induced changes of the low-energy electronic structure with unprecedented precision.

## RESULTS

Following in situ cleavage of high-purity single crystals of $Sr_3Ru_2O_7$ at low temperatures, we find large-scale atomically clean surfaces with very few defects (∼0.08% per $nm^2$). Topographic images (Fig. 1C) show atomic resolution due to the Sr lattice. In addition, at low bias voltages (Fig. 1D), a checkerboard charge order can be detected as well as a breaking of $C_4$ symmetry near defects. The checkerboard charge order is reminiscent of the one observed at the surface of $Sr_2RuO_4$ (*19*). While $Sr_3Ru_2O_7$ has an orthorhombic crystal structure (*20*), the orthorhombicity is tiny and often neglected, yet it does lead to subtle differences in the electronic structure in the crystallographic [110] and [1$\bar{1}$0] directions (*21*). Following the notation used in previous works (*1*, *10*), here, we provide crystallographic directions in the tetragonal notation (Fig. 1A). Different from the nematicity in $Sr_2RuO_4$ that occurs between [10] and [01] directions, the dominant symmetry breaking here is between the [1$\bar{1}$] and [11] directions, suggesting a different origin. Tunneling spectra (Fig. 1E) acquired at $T = 80$ mK show a pronounced gap-like structure with two peaks in the vicinity of the Fermi energy at ±5 mV. Both, topographic images and tunneling spectra, are consistent with previous reports (*22*, *23*). High-resolution tunneling spectra (inset of Fig. 1E) reveal an additional substructure of the spectra.

[1]SUPA, School of Physics and Astronomy, University of St Andrews, North Haugh, St Andrews KY16 9SS, UK. [2]School of Chemistry, University of St Andrews, North Haugh, St Andrews KY16 9ST, UK. [3]Max Planck Institute for Chemical Physics of Solids, Nöthnitzer Straße 40, 01187 Dresden, Germany.
*Corresponding author. Email: wahl@st-andrews.ac.uk.
†These authors contributed equally to this work.
‡Present address: Key Lab for Magnetism and Magnetic Materials of the Ministry of Education, Lanzhou University, Lanzhou 730000, China.

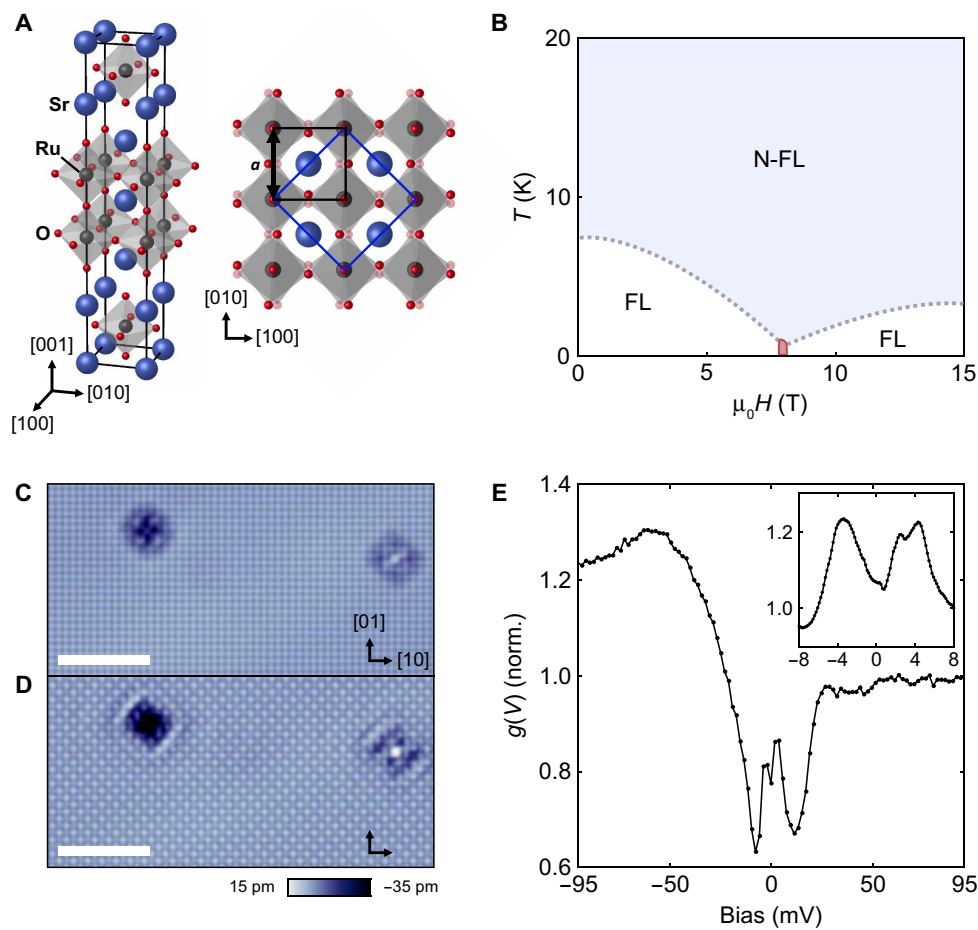

**Fig. 1. Crystal structure, phase diagram, and STM characterization of Sr$_3$Ru$_2$O$_7$.** (**A**) Bulk crystal structure with the tetragonal unit cell in side and top views. Black and blue squares indicate the projection of the tetragonal and orthorhombic unit cells, respectively. Lattice vectors are given in terms of the tetragonal unit cell (blue, Sr; black, Ru; red, O). (**B**) Sketch of the phase diagram of bulk Sr$_3$Ru$_2$O$_7$ for magnetic fields **B**||[001] based on the exponent $\alpha$ of the temperature dependence of the resistivity $\rho(T, B) = AT^{\alpha} + C$. The dotted gray lines indicate crossovers between Fermi liquid (FL, $\alpha = 2$) and non-Fermi liquid (N-FL, $\alpha \sim 1$) behavior. (**C**) Topography imaged at $V = +45$ mV (scale bar, 5 nm, $I_{set} = 818$ pA) showing the Sr lattice and two defects. (**D**) The same topographic area imaged at $V = -5$ mV (scale bar, 5 nm; $I_{set} = 91$ pA), showing a checkerboard pattern on the Sr lattice. The quasiparticle interference patterns around the defects are highly anisotropic. (**E**) Tunneling spectrum $g(V)$ measured at $T = 80$ mK ($V_{set} = 100$ mV, $I_{set} = 265$ pA, $V_L = 1.90$ mV). Inset shows a high-resolution $g(V)$ spectrum around $E_F$ [$V_{set} = 8$ mV, $I_{set} = 500$ pA, $V_L = 160$ µV, $T = 80$ mK, averaged over a (1.48-nm)$^2$ area].

Representative QPI data taken in zero field are shown in Fig. 2. The real-space images from a spectroscopic map (Fig. 2, A to E) show the wave-like patterns around defects characteristic of quasiparticle interference. Their Fourier transformations (Fig. 2, F to J) reveal three distinct sets of features: (i) a fourfold symmetric broken-up ring around **q** = (0,0), which exhibits a hole-like dispersion that terminates a few millivolts above the Fermi energy (labeled as **q**$_1$ in Fig. 2H); (ii) a set of features at **q** = (−0.25,0.25) (**q**$_2$ and **q**$_3$), which strongly break $C_4$ symmetry—their symmetry-equivalent partner around **q** = (0.25,0.25) is completely absent from the data; and (iii) pairs of horizontal and vertical lines close to (0, ± 0.5) and ( ± 0.5,0) (**q**$_4$). The additional rings around the Bragg peaks at **q** = (0.5,0.5) can be traced back to the same scattering processes as the ring around **q** = (0,0). Two additional **q** vectors are found around **q** = (0,0), which have the same magnitude at −0.3 mV (**q**$_5$ and **q**$_6$). While **q**$_5$ has a hole-like dispersion with a maximum above the Fermi level, **q**$_6$ has an electron-like dispersion along [10] and hole-like dispersion along [01], which is more clearly visible around the **q** = (0.5,0.5) peaks, with a saddle point at −4 mV (Fig. S5). We can link all these features to the Fermi surface of Sr$_3$Ru$_2$O$_7$ as obtained from angle-resolved photoemission spectroscopy (ARPES) Fig. 2K (*24*), with excellent overall agreement. Comparison of the dispersion relation with the electronic structure from a paramagnetic density functional theory (DFT) calculation suggests that the QPI shows signatures from the vHss around the M point and from the $\alpha_2$ band as indicated in Fig. 2L. Many of the **q** vectors are consistent between our measurement and the DFT calculations when assuming that they are dominated by intraband scattering and accounting for a renormalization of the band structure, with a few exceptions, such as **q**$_3$ and **q**$_5$, where, however, there are bands close by—suggesting that DFT does not faithfully capture the correlated electronic structure in all its details. The strong anisotropy of the electronic structure means that the influence of dispersion in the $z$ direction on the QPI (*25*, *26*) and tunneling spectra can here be neglected. We find effective masses between $m^* \approx 7$ and 30 for the $\gamma$ and $\alpha$ bands (see table S1), comparable to what has been estimated from ARPES (*24*) and quantum oscillations (*27*).

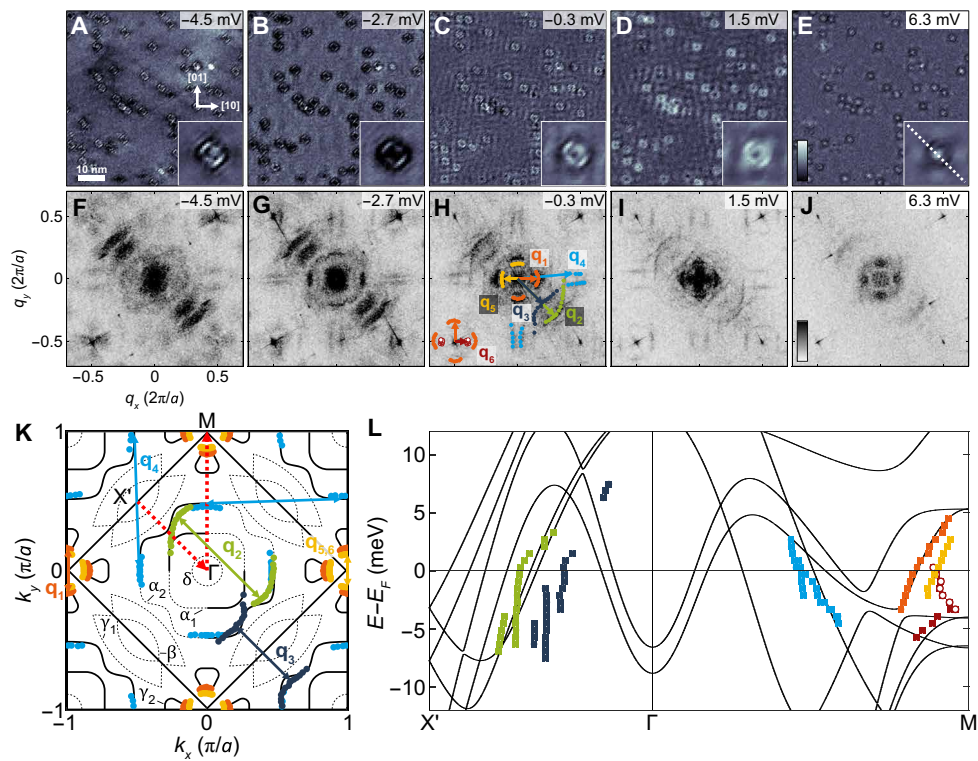

**Fig. 2. Quasiparticle interference at zero field.** (**A** to **E**) Selected real-space layers extracted from a spectroscopic map of the low energy quasiparticle interference in Sr$_3$Ru$_2$O$_7$ in an energy range of −5…6 mV ($V_{set}$ = 8 mV, $I_{set}$ = 700 pA, $V_L$ = 600 µV, $T$ = 80 mK). The insets of (A) to (E) show the average image of one type of defect with clear dispersive $C_4$ symmetry breaking features. (**F** to **J**) Fourier transforms of (A) to (E), respectively, after unfolding and correcting for drift (see fig. S4). In (H), the extracted scattering vectors $q_{1…6}$ are shown. The peaks at (± 0.5, ± 0.5) are due to the octahedral rotations and correspond to the Bragg peaks of the orthorhombic unit cell. The values of $q$ are shown in units of $2\pi/a$, where $a$ = 3.89 Å is the lattice constant of the tetragonal unit cell. (**K**) Sketch of the Fermi surface of Sr$_3$Ru$_2$O$_7$ based on the ARPES measurements of (24), overlaid with the scattering vectors shown in (H), showing very good agreement. The **q** vectors $q_5$ and $q_6$ have the same value at the Fermi level. (**L**) Comparison of the dispersion of the dominant QPI features with the band structure from a paramagnetic DFT calculation of a bilayer along the path indicated by red dotted arrows in (K). The colored points indicate the dispersions extracted from the QPI assuming the assignment shown in (H).

The most important new findings from our low-energy quasiparticle interference are (i) strong $C_4$ symmetry breaking of the QPI of the $\alpha_2$ band, suggesting that the $\alpha_2$ band is partially gapped and/or reconstructed as a consequence of a magnetic and or density wave order and (ii) the existence of a van Hove singularity above the Fermi energy, which had not been detected previously.

The detailed information revealed by low-energy quasiparticle interference allows us to extract how the electronic structure in the vicinity of the Fermi energy evolves with magnetic field. We have undertaken extensive quasiparticle interference mapping at temperatures below 100 mK as a function of field up to µ$_0H$ = 13.5 T. The appearance of the tunneling spectra changes markedly with increasing field (Fig. 3A), with substantial spectral weight shifting from above to below the Fermi energy. The change in the spectra is more complex than might be naively expected. The most prominent feature in the field dependence emerges above 10 T culminating in a peak-like feature at −1.2 mV at 13.5 T. To emphasize the field-dependent changes, we plot in Fig. 3B the spectra $\hat{g}(V, H)$ after subtracting the field-averaged background through $\hat{g}(V, H) = g(V, H) − \langle g(V, H)\rangle_H$. We can see clear evidence of a spectral feature that starts out at ~4.5 mV at 0 T and shifts to lower energy with increasing field, traversing the Fermi energy at a magnetic field of ~11 T until it reaches −1.2 mV at 13.5 T. We can correlate this behavior with field-dependent quasiparticle interference data. Figure 3C to E show the Fourier transformation of QPI maps acquired at µ$_0H$ = 0, 8 T, and 13 T with the same tip and under identical tunneling conditions. The changes in the QPI are most notable at small **q** vectors around **q** = 0. Tracing the scattering vector $q_5$ with field, we can clearly resolve the associated scattering pattern in zero magnetic field, while its magnitude is reduced at 8 T and becomes hardly visible at 13 T. This trend is more easily seen in cuts through the QPI data: Figure 3F to H show how the van Hove singularity associated with $q_5$ and with the peak in $\hat{g}(V, H)$ spectra moves across the Fermi energy with increasing field; at zero field (Fig. 3F), a clear hole-like dispersion with a band maximum at 4.5 mV is seen, consistent with the $\hat{g}(V, 0T)$ spectrum. At 8 T, the band maximum of the hole-like dispersion has shifted to ~2 mV and lastly to ~ −1 mV at 13.4 T, where it crosses the Fermi energy as an electron-like dispersion. None of the bands show evidence for spin-splitting, as would be expected if a paramagnetic band was Zeeman split in magnetic field. Our data, therefore, indicate that the surface is already magnetic.

The change of the spectra is accompanied by a qualitative change in the appearance of topographic images acquired at small bias voltages, revealing a magnetic field–induced unidirectional charge order superimposed to the checkerboard charge order. While we see strong symmetry breaking QPI around defects, topographic images acquired at zero magnetic field of defect-free areas exhibit $C_4$ symmetry (Fig. 4A); therefore, the zero field state is characterized by pure nematicity without a symmetry-breaking charge order. With an increasing magnetic field, a distinct stripe-like contrast appears in the STM images at one of the wave vectors of the orthorhombic unit cell, $\mathbf{q}_{stripe} = (0.5, 0.5)$, Fig. 4B. Occasionally, after ramping the field, this field-induced stripe order switches direction as does the nematicity of the electronic structure (Fig. 4C and D). The intensity of the field-induced stripe order, $\tilde{z}(\mathbf{q}_{stripe})$ (Fig. 4E), is directly proportional to field and becomes most prominent at the highest magnetic fields. The intensity of the second, in a tetragonal unit cell symmetry equivalent, peak at $\mathbf{q}_{stripe} = (-0.5, 0.5)$ remains constant. This is true for a wide range of bias voltages (fig. S10). The characteristic wave vector of the stripe order is always oriented in a direction normal to the dominant QPI signal associated with $\mathbf{q}_2$ and $\mathbf{q}_3$, suggesting that the $\alpha_2$ band becomes gapped out or incoherent in the direction of the stripe order.

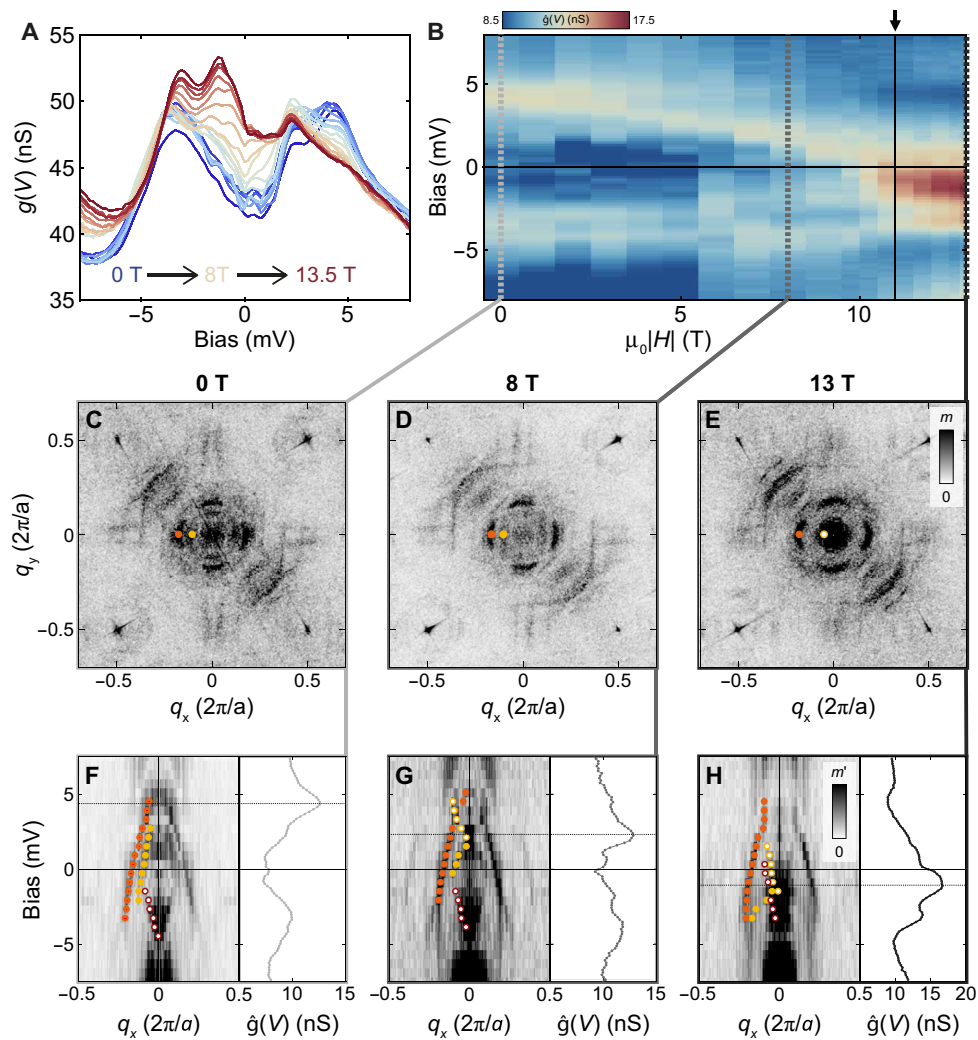

**Fig. 3. Magnetic field–induced Lifshitz transition. (A)** Tunneling spectra $g(V, H)$ as a function of field $\mu_0 H = 0 \ldots 13.4$ T. A clear spectral weight shift from positive to negative bias voltages is observed ($V_{set} = 8$ mV, $I_{set} = 500$ pA, $V_L = 160$ μV, $T = 80$ mK). **(B)** Color map plot of the differential conductance $\hat{g}(V, H)$ after subtracting the field average of the spectra, $\hat{g}(V, H) = g(V, H) - \langle g(V, H) \rangle_H$, as a function of bias $V$ and magnetic field $\mu_0 H$. The arrow indicates the field $\mu_0 H_c = 11$ T at which a peak in $\hat{g}(V, H)$ crosses the Fermi level. **(C to E)** Fourier transforms of quasiparticle interference maps at −0.3 mV for (C) $\mu_0 H = 0$ T, (D) 8 T, and (E) 13 T. Clear differences are observed around **q** = (0,0). **(F to H)** Line cuts through the spectroscopic maps along $q_x$ at (F) $\mu_0 H = 0$ T, (G) 8 T, and (H) 13 T. The circles indicate $\mathbf{q}_1$ (orange), $\mathbf{q}_5$ (yellow), and $\mathbf{q}_6$ (red), with hole-like dispersion shown as full circles and electron-like dispersion shown as open circles. The plots on the right of each line cut show the $\hat{g}(V, H)$ spectrum at each $\mu_0 H$. From (F) to (H), a vHs is seen in both the QPI and $\hat{g}(V, H)$ spectra moving from above to below the Fermi energy (indicated by horizontal dotted line).

## DISCUSSION

Our quasiparticle interference data demonstrate that the surface layer of $Sr_3Ru_2O_7$ undergoes a magnetic field–induced Lifshitz transition at a field $\mu_0 H_c = 11$ T accompanied by the emergence of a field-induced charge order. The appearance of topographic images, tunneling spectra, and the quasiparticle interference all reveal clear and substantial field-dependent changes.

The field at which we observe the Lifshitz transition in the surface layer is slightly larger than the field at which the metamagnetic critical point is observed in the bulk. We expect the small structural changes occurring in the surface layer (28), most notably an increased octahedral rotation, to lead to changes in its properties. A high sensitivity of the metamagnetic properties of $Sr_3Ru_2O_7$ to small structural changes has also been seen with uniaxial (29) and hydrostatic (30) pressure and strain (31). Although the crystal structure at the surface of $Sr_3Ru_2O_7$ is very similar to that of the surface layer of $Sr_2RuO_4$, except for the second $RuO_6$ layer, our data show different behaviors. Unlike the case of $Sr_2RuO_4$, where the $C_4$ symmetry breaking can be described by a small nematic term in a tight-binding model (19), here, we find a large $C_4$ symmetry breaking that cannot be easily captured by a small perturbative term. Although $Sr_3Ru_2O_7$ is already an orthorhombic system and therefore does not exhibit $C_4$ symmetry, the orthorhombic distortion is tiny and not expected to lead to a significant anisotropy of this scale. Furthermore, the occasional switching of the direction of the nematicity after field ramps (Fig. 4C and D) excludes a structural origin of the symmetry breaking and stripe order found here.

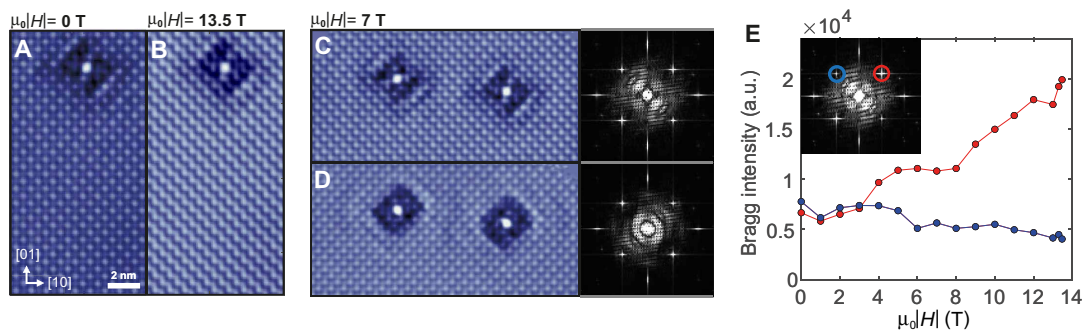

**Fig. 4. Magnetic field–induced zigzag order.** Topographic image of the same area at (**A**) $\mu_0 H = 0$ T and (**B**) $\mu_0 H = 13.5$ T. The emergence of a zigzag stripe order with magnetic field is clearly visible. (**C** and **D**) Topographic images obtained in the same area showing two neighboring defects at $\mu_0 |H| = 7$ T. Between (C) and (D), the field has been ramped, changing the orientation of the nematicity and stripes occasionally, demonstrating that the nematicity and stripe order are not of structural origin. The corresponding Fourier transforms are shown on the right-hand side. (**E**) Intensity of the Fourier peaks $\tilde{z}$(q) associated with the charge order at **q** = (0.5, −0.5) and **q** = (0.5, 0.5) as a function of magnetic field showing a strong increase in the peak intensity in one direction only, consistent with the emergence of the zigzag order. Inset: The Fourier transform $\tilde{z}$(q) of the topography taken at 13.5 T, highlighting the positions of the Fourier peaks shown in the panel. All images analyzed for (E) were taken in the same location with the same tip ($V_{set} = −5$ mV, $I_{set} = 91$ pA; a.u., arbitrary units).

Possible explanations for the twofold reconstruction of the Fermi surface that we observe include (i) the formation of a spin density wave phase or antiferromagnetic order, (ii) spin-orbit coupling induced gapping of parts of the Fermi surface, or (iii) correlation effects leading to incoherence of the bands in one direction. An $E$-type antiferromagnetic order, with a wave vector $q_{SDW}$ = (0.25,0.25), has been identified from DFT as the leading antiferromagnetic instability of $Sr_3Ru_2O_7$ (32) with only a small energy difference to the ferromagnetic ground state. Experimentally, minute amounts of impurities of Fe (33), Mn (34), or Ti (35) in $Sr_3Ru_2O_7$ result in a ground state with $E$-type order. However, the Fermi surface resulting from this ground state in DFT calculations is significantly different from the one observed here (see fig. S11), while our QPI is largely consistent with the paramagnetic electronic structure obtained from DFT and ARPES. This suggests that, if this order is realized in the surface layer, the changes in the electronic structure are not accurately captured by DFT. The magnetic order would also have a different ordering vector than the one detected in the bulk by neutron scattering (10).

Spin-orbit coupling in a polarized state has been proposed as a mechanism to explain the field-induced nematicity in bulk $Sr_3Ru_2O_7$ (16, 17). Here, this polarized state could be due to a magnetic ground state stabilized in the surface layer as suggested by our data through the absence of a splitting of bands in field.

In the magnetic-field-dependent spectroscopy close to the Fermi energy, we can see clear signatures, as also confirmed from the QPI measurements, of a vHs crossing the Fermi energy with applied magnetic field, with a slope that would correspond to a $g$ factor of ~15.

The magnetic-field dependence of the topographic images is likely a consequence of the $C_2$ symmetry of the electronic structure and the vHs crossing the Fermi energy with field: If the vHs is not pinned to the M point with fourfold symmetry, as would be the case for tetragonal symmetry, but crosses the zone boundary away from the M point, it can result in a strong appearance of the corresponding scattering vector when crossing the Fermi energy. The field-induced charge order is at a wave vector **q** = (0.5,0.5), which connects opposite edges of the Brillouin zone corresponding to the orthorhombic unit cell. We note that the recently proposed loop current state in the surface layer of $Sr_2RuO_4$ (36) would be expected to result in a similar stripe-like pattern in magnetic field. While the magnetic-field–induced charge order reported here seemingly could explain anisotropic transport signatures found near fields around the metamagnetic phase transitions in the bulk $Sr_3Ru_2O_7$ (1), it occurs in a different crystallographic direction: We observe this symmetry breaking between the [11] and [1$\bar{1}$] directions of the tetragonal unit cell, whereas the anisotropic transport and field-induced magnetic-order observed in neutron scattering are along the [10] and [01] directions (1, 10). In a scenario where the symmetry breaking is driven by spin-orbit coupling, this difference could be easily accounted for by a different easy axis direction in the surface layer compared to the bulk (37). The differences suggest that while the surface layer exhibits a very rich physics and phenomenology on par with that of the bulk, there are some subtle and not so subtle differences between the two.

Our results reveal a magnetic-field-driven Lifshitz transition accompanied by stripe order and nematicity in the surface layer of $Sr_3Ru_2O_7$, exhibiting notably similar phenomena compared to the bulk material, yet in a new regime distinct from that of the bulk. We find an intricate interplay of electronic, structural, magnetic, and orbital degrees of freedom. Our data reveal clear signatures of a vHs in QPI and tunneling spectroscopy crossing the Fermi energy at a field of $\mu_0 H = 11$ T, accompanied by field-induced stripe order. The detailed spectroscopic information about the electronic structure across the field-induced Lifshitz transition enables direct verification of microscopic models for the metamagnetic properties of $Sr_3Ru_2O_7$. The ability to spectroscopically track changes in the electronic structure across a Lifshitz transition and simultaneous imaging of emergent order associated with it promises a detailed microscopic understanding of the physics of a metamagnetic critical point.

## MATERIALS AND METHODS

### Single crystal growth

Single crystals of $Sr_3Ru_2O_7$ have been grown in a high-pressure image furnace (SciDre) at pressures of $p_{Ar,O2} = 20$ bar with a growth speed of 10.5 mm/hour using a 6500-W lamp. The crystals have been characterized by low-temperature transport measurements (see the Supplementary Materials for details) to verify their quality and that they exhibit the behavior found in high-purity crystals (*1*).

### Scanning tunneling microscopy

Experiments were performed in an ultralow-temperature STM mounted in a dilution refrigerator (*38*) and confirmed in a second STM. The energy resolution of the STM has been verified to be better than 50 μeV (*38, 39*). Tunneling spectra are recorded under open feedback loop condition using a lock-in technique, and the peak-to-peak modulation $V_L$ is specified in the image captions. Samples have been cleaved in situ at a cleaving stage at a temperature below 20 K and then directly inserted into the microscope head. Tunneling spectra as presented in the inset of Fig. 1E and in Fig. 3 (A and B) have been obtained by averaging over a 1.48-nm by 1.48-nm area corresponding to two-by-two unit cells. Additional experiments were carried out on a sample with 1% Ti-doping (see section S10 and Fig. S12).

### Transmission electron microscopy

We have performed transmission electron microscopy (TEM) to verify the structural quality of the single crystals. The sample for scanning TEM analysis was prepared by gallium focused ion beam (FIB) milling using an FEI Scios FIB scanning electron microscope equipped with an EDAX Hikari Super electron back-scattered diffraction (EBSD) detector. The orientation of the sample was determined by EBSD before milling to cut a lamella in the ⟨010⟩ plane. High-angle annular dark-field images were recorded using a probe corrected FEI Themis 200 scanning/transmission electron microscope operated at 200 kV with a probe covergence angle of 21.2 mrad and inner/outer collection angles of 88.4 and 200 mrad, respectively.

## SUPPLEMENTARY MATERIAL

Supplementary material for this article is available at https://science.org/doi/10.1126/sciadv.abo7757

**Acknowledgments:** We acknowledge useful discussions with M. P. Allan, C. A. Hooley, and A. W. Rost. **Funding:** C.A.M. acknowledges funding from EPSRC through EP/L015110/1 and LCR from the Royal Commission for the Exhibition of 1851. C.A.M., M.N., and P.W. further acknowledge funding from EPSRC through EP/R031924/1 and I.B. through the International Max Planck Research School for Chemistry and Physics of Quantum Materials. TEM measurements were supported through grants EP/R023751/1, EP/L017008/1 and EP/T019298/1 from EPSRC. **Author contributions:** C.A.M. and L.C.R. carried out STM measurements, analyzed the data, and prepared figures. C.A.M. undertook the QPI analysis. I.B. and M.N. carried out STM measurements on a second instrument, and A.B.N. did TEM measurements. L.C.R. and C.A.M. did DFT calculations. Z.L., A.C.K., and A.P.M. grew samples. P.W. initiated and led the project. P.W., L.C.R., and C.A.M. wrote the manuscript. All authors contributed to and discussed the manuscript. **Competing interests:** The authors declare that they have no competing interests. **Data and materials availability:** All data needed to evaluate the conclusions in this paper are present in the paper and/or the Supplementary Materials. Underpinning data will be made available at https://science.org/doi/10.1126/sciadv.abo7757.

Submitted 24 February 2022
Accepted 15 August 2022
Published 30 September 2022
10.1126/sciadv.abo7757

# Science Advances

## Atomic-scale imaging of emergent order at a magnetic field–induced Lifshitz transition

Carolina A. MarquesLuke C. RhodesIzidor Benedi#i#Masahiro NaritsukaAaron B. NadenZhiwei LiAlexander C. KomarekAndrew P. MackenziePeter Wahl

**View the article online**
https://www.science.org/doi/10.1126/sciadv.abo7757
**Permissions**
https://www.science.org/help/reprints-and-permissions

Use of this article is subject to the Terms of service